# Investigating Causal Cues:

# Strengthening Spoofed Audio Detection with Human-Discernible Linguistic Features

Zahra Khanjani, Tolulope Ale, Jianwu Wang, Lavon Davis, Christine Mallinson, Vandana P. Janeja

University of Maryland, Baltimore County

**Abstract:** Several types of spoofed audio, such as mimicry, replay attacks, and deepfakes, have created societal challenges to information integrity. Recently, researchers have worked with sociolinguistics experts to label spoofed audio samples with Expert Defined Linguistic Features (EDLFs) that can be discerned by the human ear: pitch, pause, word-initial and word-final release bursts of consonant stops, audible intake or outtake of breath, and overall audio quality. It is established that there is an improvement in several deepfake detection algorithms when they augmented the traditional and common features of audio data with these EDLFs. In this paper, using a hybrid dataset comprised of multiple types of spoofed audio augmented with sociolinguistic annotations, we investigate causal discovery and inferences between the discernible linguistic features and the label in the audio clips, comparing the findings of the causal models with the expert ground truth validation labeling process. Our findings suggest that the causal models indicate the utility of incorporating linguistic features to help discern spoofed audio, as well as the overall need and opportunity to incorporate human knowledge into models and techniques for strengthening AI models. The causal discovery and inference can be used as a foundation of training humans to discern spoofed audio as well as automating EDLFs labeling for the purpose of performance improvement of the common AI-based spoofed audio detectors.

## 1. Introduction

Audio spoofing involves manipulating audio signals to mimic natural human speech, via techniques such as mimicry, replay attacks, and AI-generated audio, or audio deepfakes. Perpetrators often exploit synthetic voices that sound familiar to their victims, resulting in severe consequences. For instance, in 2019, a manager of a German company transferred €220,000 to a supposed vendor based on instructions received via spoofed audio impersonating the CEO [14]. In another example, in 2021, the New York Times reported that a media company, Ozy Media, attempted a fraudulent $40 million investment from Goldman Sachs [22]. In a near final phone call, one of the voices "began to sound strange to the Goldman Sachs team, as though it might have been digitally altered." The situation turned out to be an attempt to impersonate a key executive, thwarted only by a listener's keen ear. As this situation demonstrates, fake audio poses great risks [3], [1]. Yet, there are also linguistic nuances that, if recognized, can help listeners discern it—which speaks to the untapped potential of the human listener in audio deepfake discernment. To address the challenges posed by spoofed audio, a recent study proposed augmenting the audio training data with what were called Expert Defined Linguistic Features (EDLFs) [9] as discernible features for detecting spoofed English audio. Selected by two team members who are experts in sociolinguistics, inspired by the work [9], as described more fully below, we considered five commonly occurring, predictable features of spoken English. These EDLFs were extracted from various types of spoofed audio samples in a hybrid dataset, including replay attacks, mimicry, two types of audio deepfake (text-to-speech and voice conversion), and genuine audio samples. By leveraging these discernible features, the researchers improved accuracy in classifying real and fake audio compared to a existing baseline, Linear Frequency Cepstral Coefficients (LFCCs) with a lightweight convolutional neural network (LCNN)–which was one of the best performing baselines from the recent Automatic Speaker Verification and Spoofing Countermeasures Challenge (ASVspoof 2021) [28][9]. We aimed to see if any of the EDLFs have a causal connection to the spoof label, which would indicate a stronger link for spoofed audio detection and facilitate the detection of deepfakes through better training datasets. In this study, we investigate causal relationships between the EDLFs and spoof labels to advance feature selection for the purpose of spoofed audio detection. We propose an ensemble causal discovery model to identify the EDLFs that have the most significant impact on determining the authenticity of an audio sample. This model enables us to identify causal links between the

selected EDLFs and the likelihood of audio being fake or real.

Additionally, we employ a more tightly coupled ground truth validation with the help of the sociolinguistics experts. This is a novel way of strengthening AI through human knowledge and creating a feedback loop to the models with the domain experts' input driving the algorithm refinement. The contributions of this work are:

• Causal discovery applied on the human discernible linguistic audio features for spoofed audio detection: Using labeled features extracted by sociolinguistic experts in a hybrid dataset, we find the most relevant features indicating spoof audio labels using causal discovery algorithms. We propose an ensemble method across
multiple causal models.

• Causal inference: For the EDLFs we find the estimate of average causal effect of each feature on the target variable (audio spoof), based on the given dataset and back-door criteria approach, to determine the most relevant features to the detection of spoofed audio.

• We also perform a tightly coupled ground truth validation with sociolinguistics experts. The ultimate objectives of the proposed causal analysis are as below:

• As we mentioned, in another study, the Expert Defined Linguistic Features (EDLFs), manually extracted and fed by some AI detectors, outperformed a common baseline in spoofed audio detection. This study ranks these EDLFs from a causal perspective to inform and advance EDLF-based spoofed audio detectors.

• Knowing the causal links between the EDLFs and the audio label (spoofed vs real) clarifies our steps in auto annotation of EDLFs, i.e labeling the selected EDLFs with an AI driven process along with domain validation. When we know which of a set of EDLFs have actual causal links to the audio authenticity, as well as the causal influence score associated with a given EDLF, we will know if that EDLF may be a good candidate for auto annotation with appropriate validation from domain experts. Although the benefits of the manual process have been demonstrated in another paper [9], for applying the method on a larger scale, there is a need to automate the process of labeling validated EDLFs, for which the results of this causal analysis can help.

• These human discernible EDLFs are also used in another study to train undergraduate students in spoofed audio discernment. For this purpose also the results of the causal analysis can be utilized in preparing more efficient training materials.

The aforementioned objectives are also summarized in Figure 1.

In Sections 2 and 3 of the paper, we present related work and methodology. Section 4 discusses the experimental results, and Sections 5 and 6 outline significance, impact, conclusions, and future work.

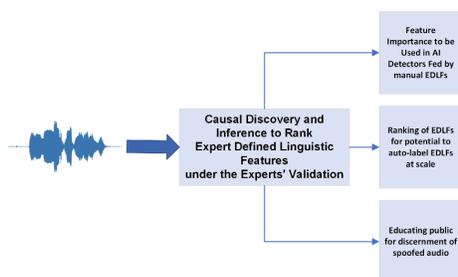

Fig. 1. Overview of the Main Objectives of the Study

## 2. Related Work

In a survey [10] of over 140 AI approaches to deepfake detection and generation in research papers, the researchers found that most studies focus on generation of video deepfakes and less on detection; of those, most detection efforts center on video and image deepfakes, and far fewer on text and audio. A recent study uses the power of layer-wise neuron activation patterns with the assumption that they can capture the imperceptible differences between fake and real audio [24]. They achieved good accuracy using a simple binary classification since their extracted features are distinguishing. However, extracting these features is complicated and requires using deep neural networks in the pre-processing phase. [15] looks at the efficiency of perceptual features such as pitch-based features in spoofed audio detection and demonstrates that those features are highly effective. In another recent work [9], they incorporated sociolinguistic knowledge to extract linguistic features, which is called Expert Defined Linguistic Features (EDLFs). The EDLFs augmented the training data, which was then fed to different machine learning models for the purpose of spoofed audio detection.

It is found that Logistic Regression classifier outperformed one of the best performing baselines, a pre-trained model of Light Convolutional Neural Network (LCNN) [25], [13] and Linear Frequency Cepstral Coefficients (LFCCs).

We discuss causal discovery as a way to further investigate the potential contribution of linguistic features to detecting spoofed audio. Causal discovery is extensively utilized in different domains such as education, healthcare, cybersecurity and more. The implications of correlation-based methods have motivated more use of causality [6]. For instance, in health care, causality is used from finding causes of infant mortality [16] to identifying the causal structure for type-2 diabetes [21], [6]. However, to the best of our knowledge, there are no studies connecting human knowledge into a causal framework for spoofed audio detection. Moreover, few studies have sought to measure and advance human deepfake detection capabilities [5].

In this study we used the human discernible linguistic features already found to be effective in spoofed English audio detection (EDLFs), investigating causal links between them and their contribution to detecting spoofed audio. To the best of our knowledge, this is the first work focusing on causal discovery and inference using linguistic features for spoofed audio detection, thus tightly coupling the human knowledge in AI models.

## 3. Methodology

The overall methodology is displayed in Figure 2. Our approach examines causal links among EDLFs and spoof labels on English audio clips. We use multiple causal models and perform an ensemble based model. We further explore causal inference to derive relationships between the EDLFs and the spoof labels, showing the contribution of the linguistic features to the spoofed nature of the audio clip. We also perform a tightly coupled sociolinguistics based ground truth validation, which validates our method and shows iterative ways to refine the augmentation of linguistic features in AI algorithms for spoofed audio and deepfake detection.

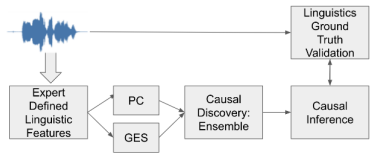

Fig. 2. Overall Methodology for investigating causal cues between human-discernible linguistic features and the spoof label

### A) Dataset

To facilitate understanding of our methodology, in this section, we briefly discuss the most commonly used English language datasets for spoofed audio detection, and explain the rationale for creating a hybrid dataset that is a combination of samples from some existing datasets and newly generated samples. We also present the properties of this hybrid dataset. Popular available spoofed audio datasets are presented in Table I. Although there are some valuable datasets for the purpose of spoofed audio detection [26], [26], [28], [20], [4], none of them have used state-of-the-art Voice Conversion algorithms such as ASSEM-VC [11], which can generate very realistic samples. Additionally, as Table I indicates, the latest datasets focus on the Text-to-Speech type of audio deepfake and ignore Voice Conversion (VC) samples [4],
[20]. Although some datasets contain VC samples, they are not labeled by their generative networks, which limits possible analysis such as [4], [20]. We need generative algorithms associated with each training audio clip for further analysis. To the best of our knowledge, for the ASVspoof 2021 dataset, this information is available only for the audio samples that are originally from ASVspoof 2019 [26] . Not only are these not new, they also have other issues.

For example, [2] examined the ASVSpoof2019 dataset for potential use for audio deepfake detection, asserting that none of the fake audio samples in the ASVspoof2019 [26] dataset are actual deepfakes. Rather, the aim of these samples is speaker verification tasks, such as voice authentication [2]. Our hybrid dataset addresses these issues. In addition to using a subset of the previous datasets [27], [20], [7], we used publicly available datasets as well as state-of-the-art generative networks, such as Melgan, Assem-VC, and Google wavenet [12], [11], [18], to create the hybrid dataset. We then incorporated sociolinguistic knowledge by adding expert identified labels for linguistic features associated with each audio clip. In our dataset, replay attack samples are from [27] with an average duration of 3.5 seconds. Text-to-Speech (TTS) samples are generated using [12], [18], and Descript samples[1], with an average duration of 4.2 seconds. Voice
Conversion (VC) samples are from [11] and their public GitHub repository, with average duration of 4.1 seconds. Mimicry samples[2] have an average duration of 2 seconds. Along with the spoofed audio samples, we include some genuine samples from the FoR dataset [20], LJ Speech dataset [7], and Obama's public speeches, with average duration of 3.8 seconds. The dataset contains 344 audio samples balanced in terms of spoofed vs genuine. The distribution of types of attacks is: 30% VC, 30% TTS, 25%replay attacks, and 15%
mimicry.

### B) Expert Defined Linguistic Features

We focus on five expert-defined linguistic features, which are called EDLFs inspired by the work [9]. The first four features are commonly occurring, variable, and distinguishing phonetic and phonological characteristics of spoken English: pitch (relative high or low tone of a speech sample), pause (break in speech production within a sample), word-initial and word-final release bursts of consonant stops (/p/, /b/, /t/, /d/, /k/, and /g/), and audible intake or outtake of breath at any point within a speech sample. As a fifth feature, we also included the experts' overall qualitative estimation of the audio quality of a speech sample. For each sound sample, two sociolinguist team members perceptually identified and annotated these features as they occurred. For the fourth feature, breath, presence or absence was captured as a binary annotation. For the remaining features, the experts captured anomalous production. For example, for a given sample, any occurrences of pitch that were perceived by the sociolinguistics experts to be anomalous in production were given a label of (PitchAnomaly =1) for that sample. Labels thus indicate potential linguistic characteristics of real versus fake audio and as such are useful tools for genuine and spoofed audio discernment. Definition 1:[Augmented Audio Clips] Given a set of n audio clips A = $a_1, a_2, \ldots a_i, \ldots, a_n$ for each audio clip $a_i$, with the Expert Defined Linguistic Features (EDLFs) such that
$a_i^{EDLF} = [a_i^{breath}, a_i^{pitch\_anomaly}, a_i^{audio\_quality\_anomaly}, a_i^{pause\_anomaly}, a_i^{burst\_anomaly}]$. Then, for the whole dataset we have:

$$A^{EDLF} = [A^{breath}, A^{pitch_{anomaly}}, A^{audio_{quality_{anomaly}}}, A^{pause_{anomaly}}, A^{burst\_anomaly}] \quad (1)$$

### C) Causal Discovery

In causal discovery our aim is to detect any causal relationships between the five EDLFs and spoofed or real audio labels. The causal discovery was geared towards identifying which linguistic features are most indicative of spoofed audio. We utilized two methods – the Constraint-based method and the Score-based method – to identify causal relationships in the EDLFs. The choice of algorithms was based on the method of edge identification in our dataset.

---
[1] https://www.descript.com/lyrebird
[2] https://www.youtube.com/watch?v=cQ54GDm1eL0

**Peter and Clark (PC) Algorithm:** PC is a constraint-based causal discovery method. It is based on the fact that under the causal Markov condition and the faithfulness assumption, when there is no latent confounder, two variables are directly causally related (with an edge in between) if and only if there is no subset of the remaining variables conditioning on which they are independent. It uses repeated conditional independence tests to determine the causal relationships among different variables in a given data set.

Definition 2: [PC Graph] Given $A^{EDLF}$, and (Label) set which includes the true labels of each $a_i$, each variable from $A^{EDLF}$ and Label represents a node (V) in the causal graph. PC Starts with a completed undirected acyclic graph where for all of the variables (V) there is an undirected edge (E) while the whole graph is acyclic.

PC applies conditional independence tests among (V) set. Then if $V_i$ is independent of $V_j$, $E_{V_i,V_j}$ will be removed. PC gets the skeleton and V-structures first, and then it orients the remaining edges.

**Greedy Equivalence Search (GES) Algorithm:** The Greedy Equivalence Search algorithm is a score-based causal discovery method. GES starts with an empty graph including the nodes as the EDLFs as variables (V), and iteratively adds directed edges such that the improvement in a model fitness measure (score) is maximized i.e. greedily adding and removing whichever valid edge most improves the score. Maximizing the score means maximizing the likelihood of observing the data with regularization on the number of parameters and the sample size.

Definition 3: [GES Graph] Given nodes V = $V_i, ..., V_n$ for each possible pair of $V_i$, and $V_j$, GES considers all possible relationships where Bayesian Information Criterion (BIC) score is calculated, the situation for which BIC is the maximum is kept in the graph, such that BIC = kln(n)−2ln(^L) here: ^L is the maximized value of the likelihood function of the model, k is the number of parameters estimated by the model, N is the number of samples in the dataset, resulting in GGES = (V, $E_{GES}$).

**Ensemble Causal Discovery:** In this method, we only keep the edges and directions that both of the aforementioned algorithms (PC and GES) agree on in terms of the existence of the edge between two nodes. Our final causal graphs are obtained from this method.

Definition 4: [Ensemble Causal Model] Given $G_{PC} = V, E_{PC}$) and, $G_{GES} = V, E_{GES}$) $G_{ensemble}$ is defined as (V, $E_{ensemble}$) where $E_{ensemble} = E_{PC} \cap E_{GES}$. In our ensemble graph, which is based on majority vote, any edges and directions that both of the two base graphs (PC and GES) agree on are kept, otherwise they are removed.

| Dataset | Details |
| --- | --- |
| ASVspoof datasets [27], [26], [28] | 2017 version: Replay attack only. 2019 version: Three separate scenarios: Spoofing att within a logical access (LA) which is generated by the latest TTS and VC technolo Replay spoofing attacks within a physical access (PA) scenario. 2021 version: (replay, and VC). In addition to the scenarios of the ASVspoof 2019, it also includes a scer called speech deepfake (DF) database. This scenario is similar to the LA task, but wit speaker verification. |
| Fake or Real Dataset (FoR)[20] | TTS samples. FoR dataset uses some high quality Text-to-Speech algorithms such as voice 3 [19] and google wavenet [18]. Contains multiple versions. The last version re-recorded version of the TTS samples. |
| Audio Deep synthesis Detection challenge (ADD) [23] | This dataset is not available even for the researchers in the field, and they only give ad permission to the challenge participants during the challenge time. |
| WaveFake [4] | TTS samples. The dataset consists of 117,985 generated audio clips (196 hours total t in both English and Japanese. |

TABLE I
SOME OF THE RECENT SPOOFED AUDIO DETECTION DATASETS

**Causal Discovery Assessment:** We conducted an assessment of the causal structural graphs generated by our ensemble model by validating and contrasting them using the sociolinguistic experts' knowledge. We then incorporated their knowledge into our causal discovery phase, and reiterated over the approved edges/directions. Also, guided by domain expert knowledge and insight, we created a ground truth validation of the causal relationships for each of the EDLFs (see section 4.4).

### D) Causal Inference

To estimate the causal effect of variable X (a given EDLF) on the target variable Y (label 'spoof or 'genuine'), we used the backdoor criterion test, an approach that enables the identification of the optimal set of variables Z (remaining EDLFs) to condition on. By conditioning on the appropriate set of variables, we can establish the true causal relationship between X and Y. The backdoor criterion test operates by strategically blocking all spurious paths that may exist between X and Y. It does so by conditioning on specific nodes in the backdoor path, excluding any nodes that are descendants of X or represent collider nodes. Conditioning on such nodes would distort the true causal association between X and Y. The causal effect of X on Y can be quantified using the adjustment formula presented in Equation 2:

$$P(Y = y|do(X = x)) = \sum_z P(Y = y|X = x, Z = z)P(Z = z) \quad (2)$$

Here, Y denotes the target variable (Label), while X is the variable for which we aim to estimate the causal effect on Label, which is one of the features from $A^{EDLF}$. Z encompasses a set of variables included in the model, and adjusting for Z allows us to isolate and measure the causal effect of X on Y. By summing over all possible values of Z, we capture the full range of potential influences from these variables on the causal relationship between X and Y. We used three different algorithms for the causal inference phase: Random Forest, Logistic Regression and XGBoost which are used often in the literature such as [29]. The aforementioned ML models are used to predict the outcome of the two scenarios: one is when the EDLF is absent (0) and the other one is when the EDLF is present (1). The difference between the average predicted outcomes is the causal effect.

## 4. EXPERIMENTAL RESULTS

In this section we include the results based on the algorithms and the hybrid dataset. We discuss the causal discovery, inference results, and tightly coupled domain expert based ground truth validation of the causal graphs.

### A. Causal Discovery

The PC algorithm generated a directed graph where the edges indicated a cause-and-effect relationship between the variables. We observed that AudioQualityAnomaly played a prominent role in the graph as it acted as a collider node for all other variables except BurstAnomaly, which had the Label variable as a descendant. This structure implied that AudioQualityAnomaly had a direct influence on the other variables, while BurstAnomaly had a direct effect only on the Label variable. There were only two direct causes for Label using PC method: BurstAnomaly and AudioQualityAnomaly. On the other hand, the GES algorithm generated an undirected acyclic graph with node connections. Unlike the PC algorithm, the GES graph did not indicate the directionality of the cause-and-effect relationships between the variables. Additionally, BurstAnomaly had no edge connections with any other variable node. To maximize the likelihood of obtaining the true causal effects of the variables on the Label variable, our ensemble model combined the agreed upon node connections from both the PC and GES models. The ensemble causal graph shown in Figures 3 and 4 and incorporated the causal structural graph of the PC model and the GES model. To validate the generated causal graph, we compared it to our ground truth knowledge provided by the sociolinguistics experts. Our validation indicated that we could not assume a direct cause-and-effect relationship among the EDLFs (as our ensemble model also indicates), but there is a possibility of statistical association. We could estimate the direct cause-and-effect relationship with our target variable (Label). Guided by the experts, we also repeated the causal discovery phase without the AudioQualityAnomaly, which captures overall audio quality broadly as opposed to the other EDLFs which are specific characteristics of spoken English. Figure 3 depicts the causal graph of the ensemble model with AudioQualityAnomaly. In this graph, AudioQualityAnomaly is treated as a front-line node with edge connections to Label, IntakeOrOuttakeofBreath, and PitchAnomaly. Figure 4 shows the ensemble causal graph including EDLFs without AudioQualityAnomaly. In this case, PitchAnomaly takes on the role of the front-line node with edge connections to Label, PauseAnomaly, and IntakeOrOuttakeofBreath. This structure when incorporating the experts' knowledge and assuming the direction from PitchAnomaly to Label indicates that PitchAnomaly has a direct impact on Label. However, the edges between PitchAnomaly with PauseAnomaly and IntakeOrOuttakeofBreath remain undirected, indicating although there is no causal relationship, there is a statistical association or correlation between them.

***Key findings***: The key findings of the causal discovery phase are as below:
When considering AudioQualityAnomaly, the only causal relationship will be from this EDLF to the authenticity of audio. Therefore, the strength of causality of AudioQualityAnomaly eliminates the causality of other EDLFs. It can be interpreted as: the AudioQualityAnomaly when exists is the dominant causal cue for spoofed audio detection. However, when the quality of a spoofed audio is normal, which human discernible linguistic features can help? For answering this question, the next finding is highlighted below.
• When we consider merely the features with linguistic nature (other EDLFs except AudioQualityAnomaly) PitchAnomaly is the only direct cause to the target variable.
• There is a statistical association between PitchAnomaly with IntakeOrOuttakeof Breath or PauseAnomaly.
• There is neither causal relationship nor statistical association between BurstAnomaly and audio authenticity.

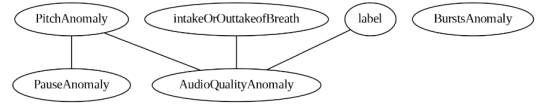

Fig. 3. Ensemble with AudioQualityAnomaly.

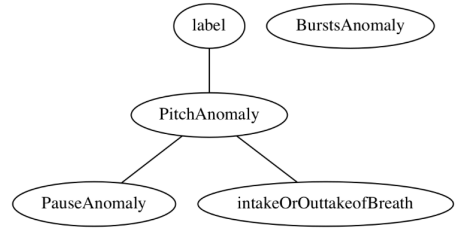

Fig. 4. Ensemble without AudioQualityAnomaly.

### B. Causal Inference

We conducted causal effect estimation on the graphs in Figure 3 and Figure 4 by conditioning on the appropriate set of variables that allowed us to obtain the causal effect using the backdoor criterion.
In Table II and Table III, we present the estimates of causal effects obtained using three different models: Logistic Regression, Random Forest Classifier, and XGBoost Classifier.
Table II displays the average causal effects without considering the AudioQualityAnomaly variable. We observed that PitchAnomaly and PauseAnomaly have the highest positive average causal effects on the Label. This indicates a strong cause-and-effect relationship between these variables. Notably, the path from PauseAnomaly to Label goes through PitchAnomaly. On the other hand, IntakeOrOuttakeofBreath exhibits an average causal effect of zero on the Label variable, although there is a path between them. This can indicate a correlation-based relationship between IntakeOrOuttakeofBreath and Label variables. These findings suggest that the outcome of the Label variable is strongly influenced by first PitchAnomaly, then PauseAnomaly. In Table III, we present the average feature, and based on that, generates the score. The sum of the scores is always 1 as they are normalized. When AudioQualityAnomaly is included, very high importance is reported for it, but very low for PitchAnomaly (the lowest score). This is because this type of feature importance is based on the specific Random Forest model. When we exclude AudioQualityAnomaly, PitchAnomaly has the most important role. Breath, consonant bursts and pause are the last layer of features respectively (PitchAnomaly > Breath > BurstAnomaly > PauseAnomaly). Although anomalous causal effects considering the presence of the AudioQualityAnomaly variable. We observed that PauseAnomaly, PitchAnomaly, and AudioQualityAnomaly have the highest causal effects on the Label variable. Furthermore, we identified a path from PauseAnomaly and PitchAnomaly through AudioQualityAnomaly leading to the Label variable. Conversely, IntakeOrOuttakeofBreath has no causal effect on the Label variable, although there is an edge between them (statistical association or correlation).
For some of the audio clips from VC and mimicry types of spoofed audio, breath does exist, thus it can lead to average causal effect of zero. These findings provide further validation of the observed average causal effects in Table II and highlight the importance of both PitchAnomaly and AudioQualityAnomaly in determining the outcome of the Label variable. These results emphasize the significance of specific variables in influencing the Label variable.
*Key Findings:* In the causal inference section, the most significant findings are as follows:
• When considering the causal influence scores using aforementioned causal inference methods, we can rank the EDLFs based on their causal effects as:
AudioQualityAnomaly > PitchAnomaly > PauseAnomaly > IntakeOrOuttakeofBreath

TABLE II
CAUSAL EFFECT OF EACH $A_{EDLF}$ (INCLUDING AUDIOQUALITYANOMALY) WITH THE TARGET VARIABLE (LABEL). LR: LOGISTIC REGRESSION, RFC: RANDOM FOREST CLASSIFIER, AND XGBC: XGBOOST CLASSIFIER

| Causal Effect Relationship | Adjustment Set | LR | RFC | XGBC |
|---|---|---|---|---|
| $PauseAnomaly \Rightarrow Label$ | $IntakeOrOuttakeofBreath$ | 1 | 1 | 1 |
| $PitchAnomaly \Rightarrow Label$ | $intakeOrOuttakeofBreath$ | 1 | 1 | 1 |
| $IntakeOrOuttakeofBreath \Rightarrow Label$ | $PauseAnomaly$ | 0 | 0 | 0 |
| $AudioQualityAnomaly \Rightarrow Label$ | { } | 1 | 1 | 1 |

TABLE III
CAUSAL EFFECT OF EACH $A_{EDLF}$ (EXCLUDING AUDIOQUALITYANOMALY) WITH THE TARGET VARIABLE (LABEL). LR: LOGISTIC REGRESSION, RFC: RANDOM FOREST CLASSIFIER, AND XGBC: XGBOOST CLASSIFIER

| Causal Effect Relationship | Adjustment Set | LR | RFC | XGBC |
|---|---|---|---|---|
| $PauseAnomaly \Rightarrow Label$ | $IntakeOrOuttakeofBreath$ | 1 | 0.8 | 1 |
| $PitchAnomaly \Rightarrow Label$ | { } | 1 | 1 | 1 |
| $IntakeOrOuttakeofBreath \Rightarrow Label$ | $PauseAnomaly$ | 0 | -0.15 | 0 |

### C. Feature Ranking
We also explored Random Forest Feature importance. Among all the trees used in random forest classifier, it calculates average of the decrease in impurity for each pitch and pause have been shown as the first and second significant features after AudioQualityAnomaly, based on both causal analysis as well as sociolinguistic experts' validation, they are not ranked properly by the Random Forest feature importance.

### D. Expert based Ground Truth Validation
Given the novel approach of embedding human knowledge in AI models we wanted to perform a tightly coupled validation of our results. Thus we present validations based on how the sociolinguistics experts view the linguistic features to understand the insights obtained from the causality analysis. In a separate process from the causality analysis, the two experts created a graphical representation of how they view the EDLFs, both in terms of their perceived relative importance as indicators of spoofed audio and the relationship of each linguistic feature to the others. The representation of these feature relationships and their importance created by the sociolinguists is provided in Figure 5.
The sociolinguistics experts' representation of the relationships and importance of the EDLFs demonstrate the utility of bringing human knowledge into AI models. Feature relationships noted by the experts aligned in several key ways with the causal graphs produced above. Also, the importance ranking produced by the experts is aligned with the causal inference results.
• First, audio quality, located within the rectangle, is represented by the sociolinguists as an overall feature that

can affect perception of the spoken features, which are represented within the circles. This assessment aligns with the AI models that suggest overall audio quality is a first line feature. Poor audio quality affects the ability to clearly and accurately hear the other four EDLFs – pitch, breath, consonant bursts, and pause. As defined for this study, audio quality is an overall indicator and is not correlated with the other four EDLFs in natural speech. The edges between audio quality and the other EDLFs can be interpreted as its effect on how we can hear the other features. In both the sociolinguists' representation and in the results of causality investigation, audio quality is demonstrated to be an encompassing cue for discerning spoofed audio. At the same time, it is important to note that not all spoofed audio has audio quality issues, and not all genuine audio has high audio quality.

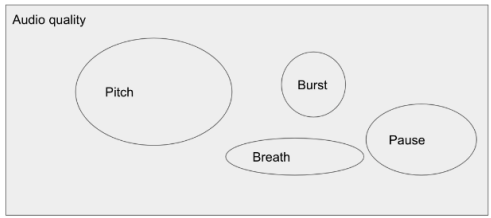

Fig. 5. Representation of Feature Relationships by Linguists

• After the first line feature of AudioQualityAnomaly, Pitch-Anomaly was the next most distinguishing feature as indicated by the causality investigation. Pitch, which is apparent in each spoken English clip, can be considered a second line feature, as suggested by the causal AI models. Similarly, pitch is represented by the sociolinguists with the largest of the four circles in the figure, indicating their assessment of its relative importance compared to the other three spoken features.

• Consonant bursts, breath, and pause, which are not necessarily present or apparent in every clip, are third level features, as suggested by the causal AI models and as similarly indicated in the sociolinguists' graphical representation. Duration of the sound clip is also an important variable for consideration, as longer durations provide more opportunities for these third line features to surface.

• Where the sociolinguists' representation of features differs somewhat from the causality analysis is with regard to BurstAnomaly. In the causality analysis, there is no edge between BurstAnomaly and Label, indicating the absence of any causal or correlation-based relationship. When the machine learning detection models were run without Burst-Anomaly, results were the same, indicating that BurstAnomaly is not a distinguishing feature for spoofed audio detection. When discussing this difference, the experts commented that they believe BurstAnomaly may still be a useful distinguishing linguistic feature, but that it may not be fully captured by the labeling method that was used. We plan to further investigate this feature going forward.

• Finally, we note that not all types of spoofed audio are equally discernable with EDLFs. In terms of types of deepfakes, mimicry is difficult to discern in general because it is a real speaker impersonating another real speaker, and therefore contains many authentic linguistic features. Voice Conversion (VC) also often contains authentic linguistic features, given how these fakes arecproduced. For mimicry especially, as well as VC in someccases, applying EDLFs to the discernment process may not work as well or as consistently as other deepfake indicators or for other types of deepfakes. Text to Speech is the easiest to capture using EDLFs because the speech is synthetically produced from text. For instance, TTS typically does not contain breaths; as such, the EDLF presence or absence of breath is a useful audio cue for discerning real versus spoofed audio for TTS. Similarly, EDLFs are useful in distinguishing sound clips generated via replay attack.

• Our sociolinguistics experts agreed that consonant burst is appropriate to exclude from the causal graphs as it may not have been best captured via their labeling.

## 5. SIGNIFICANCE AND IMPACT

Our analysis demonstrates pathways for augmenting AI models with human knowledge for better spoofed audio and audio deepfake detection. The Expert Defined Linguistic Features (EDLFs) are previously demonstrated as efficient input features in spoofed audio detector models [9]. When we investigated causal links among expert-defined linguistic features, we were able to improve the feature selection phase of spoofed audio detection. An understanding of causal links among the EDLFs further improves spoofed audio detection models by allowing for more robust feature selection and human expertise based augmentation of the training datasets. Another importance of this causal analysis is for the auto annotation and labeling of EDLFs that have been deemed worth appropriate validation from domain experts. This process augments the annotation by highlighting key portions of the audio for the experts and by including careful expert-in-the-loop training of the AI auto annotation algorithms. This process can also facilitate labeling of EDLFs at scale to utilize for the purpose of spoofed audio detection. According to our findings, based on the causal relationships, we can prioritize auto-labeling of certain features–in this case, AudioQualityAnomaly and PitchAnomaly–as a type of data augmentation to help spoofed audio detection. PauseAnomaly and IntakeOrOuttakeof Breath are the second line of features.

## 6. CONCLUSION AND FUTURE WORK

In this paper we investigated causal links between expert-defined linguistic features and spoofed audio labels, with the goal of strengthening AI algorithms for improved audio deepfake detection. This work ranks the EDLFs from a causality perspective to augment and complement the work from the domain experts. The causal discovery assists in prioritizing how human discernible linguistic features can be auto annotated along with validations from domain experts. Going forward, some researchers are also exploring auto-annotation of EDLFs internally on audio signals for future use in augmenting the feature extraction phase in a linguistically-informed manner [17]. We plan to explore variations of EDLFs that have previously been identified as well as other commonly occurring linguistic features in spoofed audio detection. Analyzing the inclusion of EDLFs in detection models from a causality perspective, once curated and vetted by experts, will also be useful in helping identify the types of spoofed audio (such as TTS and Replay attacks) for which certain linguistic features may be most promising to include.

## 7. ACKNOWLEDGMENTS

Authors would like to acknowledge support from the National Science Foundation Award #2210011. The codes and audio samples are available through our GitHub repository [8].

## REFERENCES

[1] ALMUTAIRI, Z., AND ELGIBREEN, H. A review of modern audio deepfake detection methods: Challenges and future directions. Algorithms 15, 5 (2022), 155.

[2] BLUE, L., WARREN, K., ABDULLAH, H., GIBSON, C., VARGAS, L., O'DELL, J., BUTLER, K., AND TRAYNOR, P. Who are you (i really wanna know)? detecting audio {DeepFakes} through vocal tract reconstruction. In 31st USENIX Security Symposium (USENIX Security 22) (2022), pp. 2691–2708.

[3] CHESNEY, BOBBY; CITRON, DANIELLE. Deep fakes: A looming challenge for privacy. https://lawcat.berkeley.edu/record/1136469 (2019). Publisher: California Law Review.

[4] FRANK, J., AND SCHÖNHERR, L. Wavefake: A data set to facilitate audio deepfake detection. arXiv preprint arXiv:2111.02813 (2021).

[5] GAMAGE, D., CHEN, J., GHASIYA, P., AND SASAHARA, K. Deepfakes and society: What lies ahead? In Frontiers in Fake Media Generation and Detection, Part of: Studies in automatic, Data-driven and Industrial Computation book series. Springer, 2022, pp. 3–43.

[6] HASAN, U., HOSSAIN, E., AND GANI, M. O. A survey on causal discovery methods for temporal and non-temporal data. arXiv preprint arXiv:2303.15027 (2023).

[7] ITO, K., AND JOHNSON, L. The lj speech dataset, 2017.

[8] KHANJANI, Z., ALE, T., WANG, J., AND JANEJA, V. Causalitydeepfake. https://github.com/MultiDataLab/CausalCues2023, 2023.

[9] KHANJANI, Z., DAVIS, L., TUZ, A., NWOSU, K., MALLINSON, C., AND JANEJA, V. P. Learning to listen and listening to learn: Spoofed audio detection through linguistic data augmentation. In 2023 IEEE International Conference on Intelligence and Security Informatics (ISI) (2023), IEEE, pp. 01–06.

[10] KHANJANI, Z., WATSON, G., AND JANEJA, V. P. Audio deepfakes: A survey. Frontiers in Big Data 5 (2022).

[11] KIM, K.-W., PARK, S.-W., LEE, J., AND JOE, M.-C. Assem-vc: Realistic voice conversion by assembling modern speech synthesis techniques. In 2022 IEEE International Conference on Acoustics, Speech and Signal Processing (ICASSP) (2022), pp. 6997–7001.

[12] KUMAR, K., KUMAR, R., DE BOISSIERE, T., GESTIN, L., TEOH, W. Z., SOTELO, J., DE BRÉBISSON, A., BENGIO, Y., AND COURVILLE, A. C. Melgan: Generative adversarial networks for conditional waveform synthesis. Advances in neural information processing systems 32 (2019).

[13] LAVRENTYEVA, G., NOVOSELOV, S., TSEREN, A., VOLKOVA, M., GORLANOV, A., AND KOZLOV, A. Stc antispoofing systems for the asvspoof2019 challenge. arXiv preprint arXiv:1904.05576 (2019).

[14] LEMOS, R. Ai-powered cyberattacks force change to network security. https://www.techtarget.com/searchsecurity/feature/AI-powered-cyberattacks-force-change-to-network-security (2020).

[15] LI, M., AHMADIADLI, Y., AND ZHANG, X.-P. A comparative study on physical and perceptual features for deepfake audio detection. In Proceedings of the 1st International Workshop on Deepfake Detection for Audio Multimedia (2022), pp. 35–41.

[16] MANI, S., AND COOPER, G. F. A study in causal discovery from population-based infant birth and death records. In Proceedings of the

AMIA Symposium (1999), American Medical Informatics Association, p. 315.

[17] NWOSU, K., EVERED, C., KHANJANI, Z., BHALLI, N., DAVIS, L., MALLINSON, C., AND JANEJA, V. Auto annotation of linguistic features for audio deepfake discernment. In AAAI, Assured and Trustworthy Human-centered AI (ATHAI) fall symposium (2023).

[18] OORD, A. V. D., DIELEMAN, S., ZEN, H., SIMONYAN, K., VINYALS, O., GRAVES, A., KALCHBRENNER, N., SENIOR, A., AND KAVUKCUOGLU, K. Wavenet: A generative model for raw audio. arXiv preprint arXiv:1609.03499 (2016).

[19] PING, W., PENG, K., GIBIANSKY, A., ARIK, S. O., KANNAN, A., NARANG, S., RAIMAN, J., AND MILLER, J. Deep voice 3: Scaling text-to-speech with convolutional sequence learning. arXiv preprint arXiv:1710.07654 (2017).

[20] REIMAO, R., AND TZERPOS, V. FoR: A dataset for synthetic speech detection. In 2019 International Conference on Speech Technology and Human-Computer Dialogue (SpeD) (2019), pp. 1–10.

[21] SHEN, X., MA, S., VEMURI, P., CASTRO, M. R., CARABALLO, P. J., AND SIMON, G. J. A novel method for causal structure discovery from ehr data, a demonstration on type-2 diabetes mellitus. arXiv preprint arXiv:2011.05489 (2020).

[22] SMITH, B. Goldman sachs, ozy media and a $40 million conference call gone wrong. The New York Times (2021). https://www.nytimes.com/2021/09/26/business/media/ozy-media-goldman-sachs.html, [Online; accessed January 11, 2023].

[23] TAO, J., LI, H., YI, J., FU, R., NIE, S., LIANG, S., AND WEN, Z. Add 2022: the first audio deep synthesis detection challenge. http://addchallenge.cn/add2022, 2022.

[24] WANG, R., JUEFEI-XU, F., HUANG, Y., GUO, Q., XIE, X., MA, L., AND LIU, Y. DeepSonar: Towards effective and robust detection of AI-synthesized fake voices. In Proceedings of the 28th ACM International Conference on Multimedia (2020), MM '20, Association for Computing Machinery, pp. 1207–1216.

[25] WANG, X., AND YAMAGISHI, J. A comparative study on recent neural spoofing countermeasures for synthetic speech detection. arXiv preprint arXiv:2103.11326 (2021).

[26] WANG, X., YAMAGISHI, J., TODISCO, M., DELGADO, H., NAUTSCH, A., EVANS, N., SAHIDULLAH, M., VESTMAN, V., KINNUNEN, T., LEE, K. A., ET AL. Asvspoof 2019: A large-scale public database of synthesized, converted and replayed speech. Computer Speech & Language 64 (2020), 101114.

[27] WU, Z., YAMAGISHI, J., KINNUNEN, T., HANILC¸ I, C., SAHIDULLAH, M., SIZOV, A., EVANS, N., TODISCO, M., AND DELGADO, H. Asvspoof: the automatic speaker verification spoofing and countermeasures challenge. IEEE Journal of Selected Topics in Signal Processing 11, 4 (2017), 588–604.

[28] YAMAGISHI, J., WANG, X., TODISCO, M., SAHIDULLAH, M., PATINO, J., NAUTSCH, A., LIU, X., LEE, K. A., KINNUNEN, T., EVANS, N., ET AL. Asvspoof 2021: accelerating progress in spoofed and deepfake speech detection. arXiv preprint arXiv:2109.00537 (2021).

[29] YOUMI SUK, H. K., AND KIM, J.-S. Random forests approach for causal inference with clustered observational data. Multivariate Behavioral Research 56, 6 (2021), 829–852